\documentclass{aa}
\usepackage{graphicx}
\usepackage{txfonts}
\usepackage{hyperref}
\usepackage{longtable}
\usepackage{lscape}
\usepackage{natbib}
\usepackage{color}
\bibpunct{(}{)}{;}{a}{}{,} 
\def\Msun{\hbox{$\thinspace M_{\odot}$}}
\def\Rsun{\hbox{$\thinspace R_{\odot}$}}
\def\Teff{\hbox{$\thinspace T_{\mathrm{eff}}$}}

\def\kms{\hbox{$\thinspace {\mathrm{km~s^{-1}}}$}}

\def\vt{\hbox{$\thinspace v_{\mathrm{t}}$}}

\def\feh{\hbox{$\thinspace [\mathrm{Fe/H}]$}}

\begin{document}

   \title{The Penn State - Toru\'n Centre for Astronomy Planet Search stars IV. Dwarfs and the complete sample
    \thanks{Based on observations obtained with the Hobby-Eberly Telescope, 
    which is a joint project of the University of Texas at Austin, Pennsylvania State University, Stanford University, 
    Ludwig-Maximilians-Universit\"at M\"unchen, and Georg-August-Universit\"at 
    G\"ottingen.}
    \thanks{Tables 3 -5 are only available online at CDS.}
        }

   
   \titlerunning{PTPS stars. IV. Dwarfs and the complete sample}
   \authorrunning{Deka-Szymankiewicz, Niedzielski et al.}

   \author{
           B. Deka-Szymankiewicz
          \inst{1}
                    \and
                    A. Niedzielski
          \inst{1}
          \and
           M. Adamczyk
          \inst{1}
         \and
           M. Adam\'ow
          \inst{2,1}
          \and
          G. Nowak
          \inst{3,4}
          \and          
          A. Wolszczan
          \inst{5,6} 
          }

   \institute{Toru\'n Centre for Astronomy, Nicolaus Copernicus University 
   in Toru\'n, Grudziadzka 5, 87-100 Toru\'n, Poland\\
              \email{bdeka@doktorant.umk.pl,Andrzej.Niedzielski@umk.pl}
            \and 
            McDonald Observatory and Department of Astronomy, University of Texas at Austin, 
            2515 Speedway, Stop C1402, Austin, Texas, 78712-1206, USA
            \and
            Instituto de Astrof\'{\i}sica de Canarias, C/ V\'{\i}a L\'actea, s/n, E38205 - La Laguna,Tenerife, Spain
            \and
            Departamento de Astrof\'{\i}sica, Universidad de La Laguna, E-38206 La Laguna, Tenerife, Spain
          \and
             Department of Astronomy and Astrophysics, 
             Pennsylvania State University, 525 Davey Laboratory, 
             University Park, PA 16802
         \and
             Center for Exoplanets and Habitable Worlds, 
             Pennsylvania State University, 525 Davey Laboratory, 
             University Park, PA 16802           
             }

   \date{Received ; accepted }

 
  \abstract
{Our knowledge of the intrinsic parameters of exoplanets is as precise as our determinations of their stellar hosts parameters. In the case of radial velocity  searches for planets, stellar masses appear to be crucial. But before estimating stellar masses properly, detailed spectroscopic analysis is essential. With this paper we conclude a general spectroscopic description of the Pennsylvania-Toru\'n Planet Search (PTPS) sample of stars.}
{We aim at a detailed description of basic parameters of stars representing the complete PTPS sample. We present atmospheric and physical parameters for dwarf stars observed within the PTPS along with updated physical parameters for the remaining stars from this sample after the first Gaia data release.}
{We used high resolution (R=60 000) and high signal-to-noise-ratio (S/N=150-250) spectra from the Hobby-Eberly Telescope and its High Resolution Spectrograph. Stellar atmospheric parameters  were determined through a strictly 
spectroscopic local thermodynamic equilibrium analysis (LTE) of the equivalent widths of Fe\,I and Fe\,II lines. 
Stellar masses, ages,  and luminosities  were estimated through a Bayesian analysis of 
theoretical isochrones. }
{ We present $\Teff$, $\log g$, [Fe/H], micrturbulence velocities, absolute radial velocities, and rotational velocities for 156 stars from the dwarf sample of PTPS. For most of these stars these are the first determinations. 
We  refine the definition of PTPS subsamples of stars (giants, subgiants, and dwarfs) and update  the luminosity classes for all PTPS stars. Using available Gaia and Hipparcos parallaxes, we redetermine the stellar parameters (masses, radii, luminosities, and ages) for 451 PTPS stars. }
{ The complete  PTPS sample of 885 stars is composed of 132 dwarfs, 238 subgiants, and 515 giants, of which the vast majority are of  roughly solar mass; however, 114 have masses higher than 1.5 M${_\odot}$ and 30 of over 2 M${_\odot}$. The PTPS extends toward much less metal abundant and much more distant stars than other planet search projects aimed at detecting planets around evolved stars;  29$\%$ of our targets belong to the Galactic thick disc and 2$\% $ belong to the halo.}


   \maketitle
%

\section{Introduction}

After the pioneering discoveries of exoplanets by 
\cite{Wolszczan1992,Mayor1995,Marcy1996}, over 3500 more planets around other stars have been identified (http://exoplanets.eu/ \citealt{Schneider2011}). { Given the limitations of current planet detection techniques and the general picture of planet formation \citep{Pollack1996}, it is safe to assume that planets are common and that, at some stage of their
evolution, many stars host or have hosted   planets born around them. }

 Therefore, studies of planetary systems at various stages of stellar evolution are of special interest. Most known exoplanets orbit main sequence  (MS) stars that are of roughly solar mass, and only a handful of projects are devoted to search for planet around stars beyond the MS, preferentially more massive than the Sun.

Intermediate-mass MS stars have high effective temperatures and rotate rapidly. The low number of spectral lines available in their spectra make these stars unsuitable for high-precision radial velocity (RV) searches for planetary companions. { Unfortunately, also in transit searches (cf. \cite{Borucki2011,Schwamb2013}) planetary candidates around such stars are discovered only occasionally,  
and 
these projects have delivered very few planetary systems around stars much more massive than the Sun  (for instance Kepler-432; \citealt{Ciceri2015,Ortiz2015,Quinn2015} or Kepler-435 \citealt{Almenara2015}). }

Of the two most efficient techniques {for searching for planets,} precise RV measurements and stellar transits, the latter has certainly delivered most of data. However in the search for planets around more massive, evolved stars the RV technique proves to be more useful. 

The RV searches that focus on stars past the MS phase (giant and subgiant stars)  study {stars} that have considerably slowed down  their rotation and have lowered their effective temperatures in terms of stellar evolution. These stars exhibit an abundant narrow-line spectrum that makes them accessible to the RV technique. Obviously, planetary systems around these stars may have already been altered by stellar evolution or dynamical interactions with other bodies during the billions of years that have passed since their formation.

The combined result of intense research in projects such as the McDonald Observatory Planet Search \citep{Cochran1993,Hatzes1993}, Okayama Planet Search \citep{Sato2005}, Tautenberg Planet Search \citep{Hatzes2005}, Lick K-giant Survey \citep{Frink2002}, ESO FEROS planet search \citep{Setiawan2003b, Setiawan2003a}, Retired A Stars and Their Companions \citep{Johnson2007}, Coralie \& HARPS search \citep{Lovis2007}, the { Bohyunsan} Planet Search \citep{Lee2011}, and our own Pennsylvania-Toru\'n Planet Search (PTPS) \citep{Niedzielski2007,Niedzielski2008}
is a  population of 103 substellar companions in 94 systems\footnote{https://www.lsw.uni-heidelberg.de/users/sreffert/giantplanets/giantplanets.php} around giant stars. 

These projects have already delivered very interesting discoveries of planets around { intermediate-mass stars    
(for instance HD 13189 \citep{Hatzes2005},   $M/M_{\odot} =  4.5 \pm2.5 M_{\odot}$;
o  UMa \citep{Sato2012}, $M/M_{\odot} =3.09 \pm0.07 M_{\odot}$)}
and reported  
 a paucity of planets within 0.5 AU of {giant  stars} \citep{Johnson2007,Sato2008,Jones2014} with only very few exceptions. 
  One of the most intriguing exceptions to that rule is  Kepler 91 b \citep{Lillo2014, Barclay2015}, a planet that orbits  a giant star in the most tight orbit of only about 2.5 R$_{\star}$.
 Our own project  already delivered  21 giants with planets, including evidence for recent violent star-planet interactions in ageing planetary systems of BD+48 740 \citep{Adamow2012}, or a rare transition object, a warm Jupiter TYC 3667-1280-1 b \citep{tapas4} orbiting a giant star.
 
  In addition to planet discoveries these surveys aim at delivering more general properties of evolved stars hosting planets.
 Relations between planet occurrence rate and stellar properties, such as stellar mass or metallicity, are of special interest. The planet occurrence rate versus stellar metallicity relation  \citep{Fischer2005,Udry2007} is well established for MS planetary hosts and giant planets. In the case of evolved stars it remains a question of a debate  (compare \cite{Pasquini2007, Zielinski2010, Takeda2008, Mortier2013, Maldonado2013} and \cite{Hekker2007,Reffert2015,Jones2016,Wittenmyer2017}).

The PTPS is a project devoted to detection and characterisation of planetary systems around stars at various stages of stellar evolution, but mainly including those more evolved than the Sun. 
With this paper we continue a series devoted to detailed  characterisation of the PTPS sample of stars. In \cite{Zielinski2012} (Paper I) we presented spectroscopic analysis of 348 red giants, mostly from the red giant clump. In \cite{Adamow2014}  (Paper II)  Li and $\alpha$-elements abundances, as well as rotational velocities for 348 stars from Paper I were presented. In a side paper, \cite{Adamczyk2016} presented a new approach to derive stellar masses and luminosities, and updated  masses, luminosities, ages, and radii for 342 stars for stars from Paper I. In the following paper, \cite{Niedzielski2016a} (Paper III)  spectroscopic analysis of additional 455 stars (subgiants and giants) was presented (see corrigendum for Paper III).

In the present paper, we deliver a spectroscopic analysis of the last sample of dwarf stars and updated stellar parameters for the complete PTPS sample after the first release of Gaia  \citep{Gaia2016} parallaxes.
We also present 
a new definition of PTPS samples of stars at various stages of evolution: dwarfs, subgiants, and giants. Consequently we present the complete PTPS sample of stars, ranging from the main sequence, through subgiant and giant branches and up to the horizontal branch (red giant clump), for which basic atmospheric and physical parameters are determined in a uniform way.

 The paper is organised as follows: The observational material and sample are presented in Section 2. In Section 3 we describe the spectroscopic analysis of our data. Section 4 contains a short presentation of the methodology used to derive the following stellar parameters: luminosities, masses, ages, and radii. In Section 5 we describe 
the method of luminosity class (evolutionary stage) assessment  and  a new subsample definition. Distribution of PTPS stars in the Galaxy is presented in Section 6. 
A summary of atmospheric and stellar parameters for the complete PTPS  sample is presented in Section 7.  
Section 8 contains a short  summary and conclusions of this paper. 

\begin{figure}
   \centering
   \includegraphics{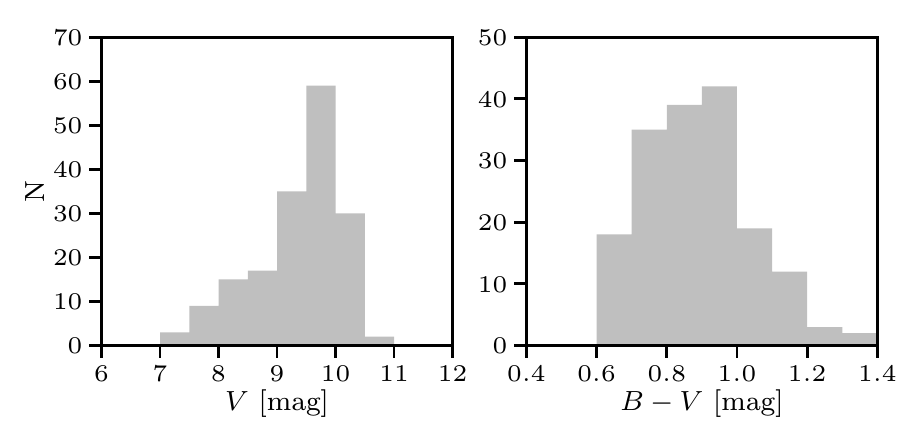}
      \caption{Histograms of the apparent magnitudes in the $V$ band (left panel) 
                   and $(B-V)$ (right panel) for 170  dwarfs for which the spectroscopic analysis was completed. }
   \label{fig-obs}
\end{figure}

\begin{figure}
   \centering
   \includegraphics{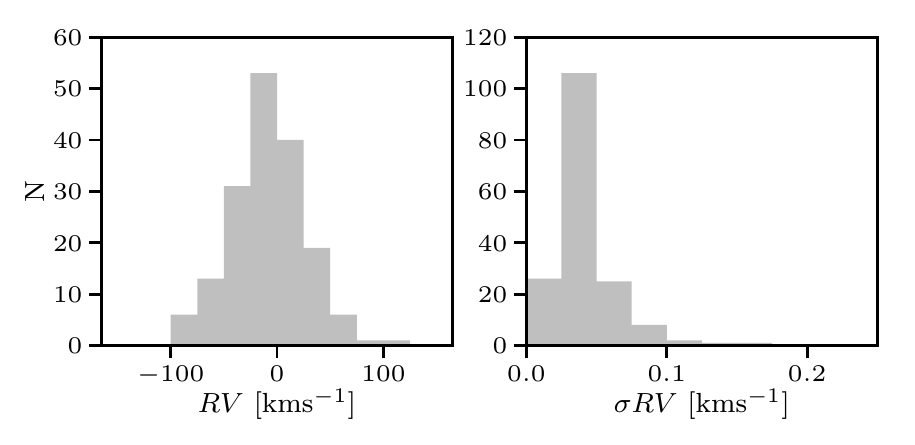}
      \caption{Histograms of the absolute RVs obtained from the cross-correlation 
                    function (left panel) and their { formal} uncertainties (right panel) for  170 PTPS dwarf
                     stars studied in this paper (Section \ref{abs-RV}).}
   \label{fig-rvcomp}
\end{figure}

\section{Targets selection and observations}\label{targets}

The sample of stars presented in this paper is the last subsample of PTPS that has not yet been presented in detail. This sample is composed of (originally) 221 dwarf stars that were selected from the compilation of Tycho \citep{vanLeeuwen2007} and 2MASS \citep{Cutri2003} catalogues by \citep{Gelino2005}. { The sample contains stars presumably from the
MS of stellar types from G to early M. The stars are expected to be field dwarfs that are {{randomly distributed across the sky accessible with the Hobby-Eberly Telescope (HET)}}, with apparent magnitudes (V) between  7.5 and 10.5 mag, and B-V colour index ($B-V$) { between  0.6 and   1.3}, approximately. The histogram of $V$ and $B-V$ for dwarfs adopted for spectroscopic analysis   
is presented in Figure \ref{fig-obs}.

Spectroscopic observations presented here were made  with the HET \citep{Ramsey1998}  and its High Resolution Spectrograph (HRS, \citealt{Tull1998}) in the queue scheduled mode \citep{Shetrone2007}. 
The spectrograph was used in the R=60~000 resolution 
and it was fed with a 2~arcsec fibre. 
The instrumental configuration and observing procedure were identical to those described 
in Paper I and Paper III. 

The collected spectra consist of 46 blue echelle orders (407 - 592~nm) and 24 
red orders (602 - 784~nm). Data reduction was carried out with a pipeline based on
IRAF\footnote{IRAF is distributed by the National Optical Astronomy 
Observatories, which are operated by the Association of Universities for 
Research in Astronomy, Inc., under cooperative agreement with the National 
Science Foundation.} tasks (flat fielding, wavelength calibration, and normalisation to 
continuum). The S/N was typically better than 200 per resolution 
element. For every star at least one so-called GC0 spectrum,  a spectrum obtained without a I$_2$ gas cell 
inserted into optical path, and a series of GC1 spectra, obtained with the gas cell inserted,  were available.
All stars studied in this paper, as well as all other PTPS targets with updated parameters, are listed in Tables 3 - 5 \footnote{Tables 3 -5 are only available  at the CDS in an electronic form.}.

\section{Spectroscopic analysis}\label{s_analysis}

The  analysis that we completed in the presented paper was a three-tier process. The aim was (1) to obtain spectroscopic parameters for the dwarfs presented here, but also (2) to reorganise the complete PTPS sample,  in a
uniform way, into subsamples based on the three luminosity classes  dwarfs, subgiants, and giants.{  Furthermore, (3) stellar parameters for all the PTPS stars were re-determined, when possible, via the application of the new Gaia   \citep{Gaia2016}  parallaxes and the already published (Paper I and III) atmospheric parameters.} 

Before that analysis, however,  we first checked all stars for spectral variability by inspecting the cross-correlation functions (CCF), following the procedure described in detail in Paper III (Section \ref{CCF}). {This step allowed us to reject from the sample all spectroscopically variable objects, such as SB2 stars, and  weak line objects that were unsuitable for the precise measurement of equivalent widths (EWs) 
and  RVs}. The process of cleaning the sample was continued  after spectroscopic parameters were obtained.  We rejected objects for which inconsistent parameters were derived from the sample  (Section \ref{ATPar}). Finally, once we estimated  stellar parameters and assigned luminosity classes, we rejected all objects with contradicting parameters  from the
sample (Section \ref{PTPS_sample}).

\subsection{CCF analysis}\label{CCF}
 
Following the results of spectral variability analysis of the subgiants sample, presented in Paper III, we performed an identical analysis of the dwarfs presented in this paper, and for all stars from  Paper I.  This way the  CCF analysis was completed for the whole sample of PTPS stars. 

After cleaning all available GC1 stellar spectra (first 17 orders) for every star  from the I2 lines, using the ALICE code \citep{Nowak2012, Nowak2013},  we cross-correlated 
 the spectra with a numerical mask consisting of 1 and 0 
value points.  
The non-zero points correspond to the positions of 
300 non-blended, isolated stellar absorption lines at zero velocity, which are present in 
a synthetic ATLAS9 \citep{Kurucz1993} spectrum of a K2 star.  
The CCF was computed step by step for each 
velocity point in a single spectral order. 

The CCFs from all spectral orders were finally added to obtain the final CCF for the whole 
spectrum.  A more detailed description of the line profile variability types identified with the CCF is presented in Paper III.

The shape of the CCF and its variation in the series of available spectra were used to identify spectroscopic 
binaries with resolved spectral line systems (SB2), objects with variable CCF  (unresolved spectroscopic binaries), and 
stars  with  flat CCF (fast rotators or low metallicity stars). 

The CCF analysis revealed that 31 stars in the dwarf sample are actually SB2 and 20 present weak or variable CCF. All of these stars are not suitable for detailed spectroscopic analysis with the applied methodology and were rejected from the sample of stars studied in more detail. Thus the dwarf sample was reduced to 170 stars. 

Similar analysis was also performed for all stars presented in Paper I. 

In total 111 stars were rejected from the PTPS  sample based on results of CCF analysis, including those  already identified in  Paper  III.
A summary of CCF analysis in the complete PTPS sample is presented in Table 3. The table contains the star identification, $\Teff$ and $\log g$ obtained from empirical calibrations (see below), with uncertainty estimates, and absolute RV associated with  the epoch of observation.

 As for those rejected stars our detailed spectroscopic analysis was not applicable, and  
 we adopted the results presented by \citet{Adamow2010}  for these stars to give an estimate of their atmospheric parameters.

In \citet{Adamow2010} the first, rough estimates of atmospheric parameters for all  stars observed within PTPS were presented. The $\Teff$ were obtained from empirical calibration of 
\citet{Ramirez2005} based on Tycho and 2MASS photometry. 
Next,  the luminosity class was estimated using two methods: reduced proper motion (RPM; \citealt{Gelino2005}) and V versus 2MASS magnitudes classification \citep{Bilir2006}. 
The RPM is an analogue of absolute magnitude. Its calculation is based on the assumption that nearby stars have high proper motions, while stars at large distances move slowly, projected in the sky. 
The colour-RPM diagram is a case of Hertzsprung-Russell diagram (HRD), hence it allows us to attribute an evolutionary status to a star with unknown distance/parallax.
Although the assumption of the distance - proper motion relation for stars is generally correct, the method does not apply to objects with atypical, that is very high or low, proper motions, which may be classified incorrectly.  With the  effective temperature and luminosity class evaluated, $\log g$ were estimated using calibrations by 
\citet{1981Ap&SS..80..353S}.

The results of this simplified analysis for 111 stars rejected from the PTPS sample are presented in Table 3.

\begin{table*}
\begin{footnotesize}
\begin{center}
\caption{\label{table:arc} Atmospheric parameters of Arcturus}
\begin{tabular}{lcccc}
 \\ \hline
References &  $\Teff$ [K] & $l\log g$  & $\vt$ [km/s] & $\feh$\\
\hline
 This work & 4235 $\pm$ 15 & 1.59 $\pm$ 0.06 & 1.35 $\pm$ 0.04 &  -0.61 $\pm$ 0.02\\
Paper III & 4254 $\pm$ 20 & 1.61 $\pm$ 0.08 & 1.54 $\pm$ 0.07 & -0.61 $\pm$ 0.08\\
\cite{Ramarc2011} &  4286 $\pm$ 30 & 1.66 $\pm$ 0.5 & & -0.52 $\pm$ 0.04\\
\cite{Mararc1977}  & 4300 $\pm$ 90 & 1.74 $\pm$ 0.2 & 1.70  &  -0.51 $\pm$ 0.08\\

\hline
\end{tabular}
\end{center}
\end{footnotesize}
\end{table*}

\subsection{Atmospheric parameters}\label{ATPar}

Stars that were classified through our CCF analysis as single or SB1 (170) were further studied in more detail.

Stellar atmospheric parameters {  (effective temperature, gravitational acceleration at the stellar
surface, metallicity ([Fe/H]), and microturbulence velocity)} were determined with the TGVIT \citep{Takeda2005}  following the procedure described in  detail in Paper I and Paper III. These papers used the line list of Paper I.

The only difference in the approach adopted in this paper is that instead of   DAOSPEC \citep{Stetson2008}, which was used in the  giant sample in  Paper I, and ARES \citep{1}, which was used for  the sample of subgiants (Paper III), we used
a new tool, 
 pyEW\footnote{https://github.com/madamow/pyEW} \citep{Adamow2015}, to measure the EWs of spectral lines. The   
 pyEW tool works on the same principle as ARES or ARES2 \citep{2}. This tool is a set of Python functions that 
are used to measure the EW of spectral lines using a Gaussian or Lorenzian function. 
To  verify results from  pyEW we compared it with  ARES. We compared the EW of iron lines in use for three stars and  found{ that they agree very well { within 0.329 m$\AA$ on average with 
 EW$_{ARES}$ / EW$_{pyEW}$ =1.011

(Pearson correlation coefficient of $r$=0.955).  }

To check for consistency with results presented in Paper III, we determined Arcturus atmospheric parameters with the adopted procedure. This star is an excellent example of a well-studied giant. We used the best quality spectrum of Arcturus from our repository,  obtained EWs of  FeI and FeII
with pyEW, and subsequently atmospheric parameters with TGVIT. Results are presented in  Table \ref{table:arc}. One can see an agreement between our atmospheric parameters and previous determination  within 1$\sigma$, except $\vt$, for which results agree within 3$\sigma$. We also note the agreement between our results and those by other authors.

Results of our determinations of atmospheric parameters for 
dwarfs are presented in Tables 4 and 5. 
For completeness in Tables 4  we also placed all other PTPS stars, for which atmospheric parameters were  presented already in Paper I and Paper III,
representing together the complete sample. A discussion of the obtained results is presented in Section \ref{PTPS_complete_sample}.

\subsubsection{Atmospheric parameters uncertainty estimates \label{errors}}

The atmospheric parameters uncertainties as delivered by TGVIT \citep{Takeda2005} represent more the quality of a fit than a real uncertainty in a given physical parameter. Only in the case of $\feh $ alone, where the uncertainty is calculated from the actual data point spread it has a regular meaning. In the case of $\Teff$, $\log g,$ and $\vt$, the uncertainties presented in Table 4 and 5 should be multiplied by a factor of 3 as discussed in detail in Paper I. We present the original uncertainties delivered by TGVIT for consistency.

\subsubsection{Comparison with literature }

A comparison of effective temperatures for 21 dwarfs from our sample, for which results of other estimates are available from Pastel Catalogue \citep{PASTEL},
is presented in Figure \ref{fig:Pastel}. A very good agreement is apparent. Unfortunately a systematic comparison of gravitational accelerations or metallicities is not possible, as these are available in the literature for only five stars.

   \begin{figure*}
\centering
 \includegraphics{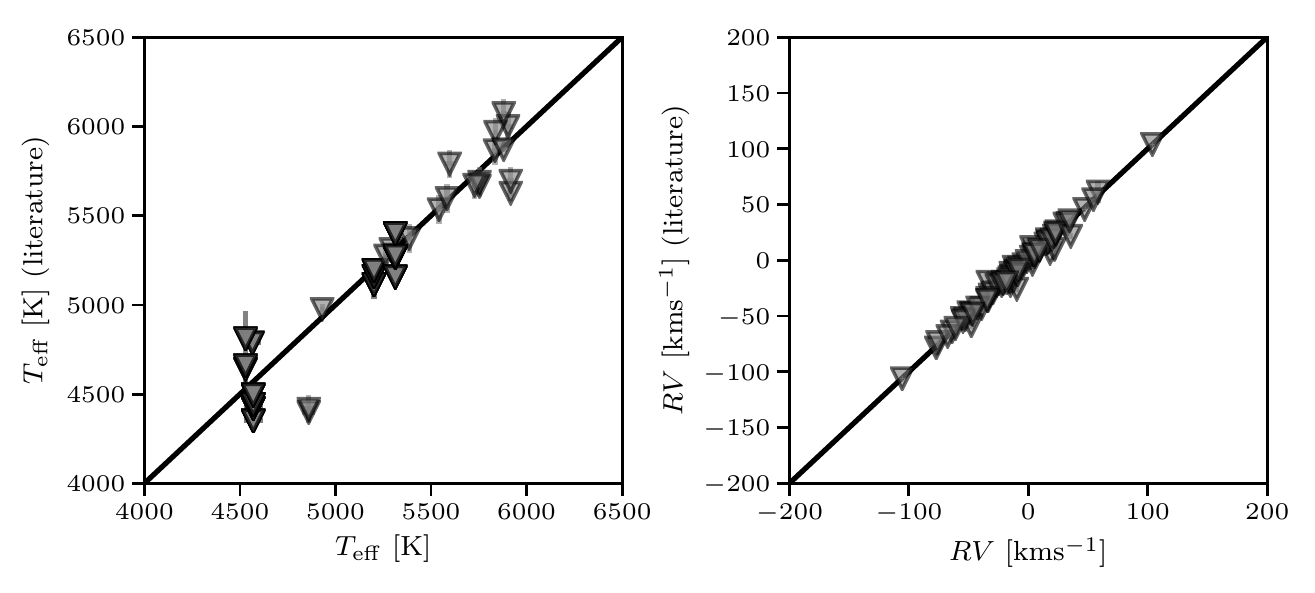}
\caption{Comparison of T$_{eff}$ and absolute RV for 21 stars in common with Pastel Catalogue \citep{PASTEL}.
}

\label{fig:Pastel}
\end{figure*}

\subsection{Absolute radial velocities}\label{abs-RV}

 The absolute RV for all stars observed within PTPS  were measured by  \cite{Nowak2012} 
with the ALICE code \citep{Nowak2012, Nowak2013} 
by fitting  a Gaussian function to 
the co-added CCFs for the 17 orders of the blue spectrum of the best available 
GC0 spectrum. 
 The RVs were transformed to the barycentre of the 
solar system with the algorithm of \citet{Stumpff1980}. 

The { formal} uncertainties of the resulting RVs were computed as rms/$\sqrt{17}$ of the 17 RVs obtained for each order separately.
The mean standard uncertainty obtained in this way is 
$\sigma \rm{RV}_{CCF}=0.039 \kms$. { However, we stress that the accuracy of the absolute RV determination is lower than the formal $\sigma \rm{RV}_{CCF}$.} As 
HET/HRS is neither thermally nor pressure stabilised, the RVs
are subject to seasonal variations and the actual precision is { in the $ \kms$ range (cf. Paper I, Paper III)}. The 
distribution of  absolute RVs for dwarfs and their uncertainties is presented in Fig.~\ref{fig-rvcomp}.

The absolute RVs with the formal uncertainties and  epochs of the observations  are presented in Tables 3, 4 and 5.

\subsection{Rotational velocities}

To obtain an estimate of  the projected rotational velocities, $v \sin i_{\star}$, 
we modelled the available GC0 and red stellar spectra with the Spectroscopy Made Easy tool (SME;  \citealt{ValPisk1996}). 
A more detailed description of this method is presented in Paper II.
The majority of our objects are slow rotators with $v \sin i_{\star}$ of $1 -3 \kms$ (Tables 4 and 5).

\section{Luminosities, masses, ages, and radii}\label{LMAR}

We estimated the stellar parameters using the Bayesian approach of  \cite{Jorgensen2005}, modified by \cite{daSilva2006} and adopted for our project by \cite{Adamczyk2016}. 
We chose  the theoretical stellar models from Bressan et al. (2012) from which we selected isochrones with metallicity Z = 0.0001, 0.0004, 0.0008, 0.001, 0.002, 0.004, 0.006, 0.008, 0.01, 0.0152, 0.02, 0.025, 0.03, 0.04, 0.05, 0.06, and 0.008 interval in log (age/yr). The adopted solar distribution of heavy elements corresponds to the solar metallicity: Z ≃ 0.0152 (Caffau et al. 2011). The helium abundance for a given metallicity was obtained from the relation Y = 0.2485 + 1.78 Z.  

For a given star, represented by  the available atmospheric parameters (together with the luminosity, if the parallax was available), 
and an isochrone of  some [Fe/H] and age t, we calculated the probability of belonging to a given mass range. The procedure used was detailed in da Silva et al. (2006). Following these authors we further calculated the searched quantities (e.g. mass, luminosity, and age) and their uncertainties from the basic parameters (mean, variance, etc.) of the normalised probability distribution functions (PDFs). 
This procedure  was described in detail  in  \cite{Adamczyk2016} and in Paper III.

For stars with known Gaia  \citep{Gaia2016}  or Hipparcos \citep{vanLeeuwen2007} parallaxes, such as $\sigma_{\pi} < \pi$ and $\pi >$ 3 mas, the luminosity was calculated directly from the visual brightness and $B-V$. For the interstellar reddening \citep{Schild1977} determination, the intrinsic $B-V$ was calculated from the effective temperature, according to the calibration of \cite{Ramirez2005}.  There are 672 PTPS objects in the Gaia catalogue, including 339 stars with previous  Hipparcos parallaxes.

 In the case in which no parallax was available or the parallax was smaller or equal 3.0 mas, the stellar luminosity was estimated from the Bayesian analysis, together with the mass and age;  
 {the limiting value of parallaxes was applied as a result of numerous calculations, which showed that the accuracy of determined physical parameters decreases rapidly with $\pi{\leq}3.0$. }
  Additionally, using the luminosity obtained from the Bayesian method we determined a Bayesian parallax ($\pi_{B}$) that was also used in the distance estimates.

As in the two previous papers  we calculated the stellar radii with two methods. First, from the effective temperature and the luminosity,

\begin{equation}
\label{eq:r1}
 R/\Rsun(\Teff,L) = \Bigg( \frac{L}{ L_{\odot}} \Bigg)^{1/2} \Bigg( \frac{\Teff_{\odot}}{\Teff}\Bigg)^{2}
,\end{equation} 
where $\Rsun$ = 695 700 km, $\Teff_{\odot}$ = 5772 K, and  $L_{\odot}$ = 3.83 $\cdot$ 10$^{26}$ \citep{Prsa2016}.

The second determination requires  $\log g$ and the mass of a star as follows:

\begin{equation}
 \label{eq:r2}
R/\Rsun(g,M)= \Bigg( \frac{M}{\Msun} \frac{g_{\odot}}{g} \Bigg)^{1/2}
\end{equation}
where $\Msun$ = 1.98855 $\cdot$ $10^{30}$ kg and $\log g_{\odot}$  = 4.44 \citep{Prsa2016}.

The average value of those two values was adopted as the final stellar radius.

For 451 stars  (including 130 stars from the presented dwarf sample) luminosities  were calculated based on new Gaia \citep{Gaia2016} parallaxes, and only masses, ages, and radii were obtained from the Bayesian analysis.
For 136 stars (including 12 dwarfs from presented sample) without Gaia parallax stellar parameters were obtained using Hipparcos parallax. 

For 298 stars (14 from the dwarf sample) with no parallaxes available, or parallaxes lower than 3 mas luminosities, masses, ages and radii were obtained from the Bayesian analysis. 

The main factor determining the precision of the stellar luminosities is the precision of the available parallaxes (587 stars). In the case of stars with unknown parallaxes (298 stars) we can only use the Bayesian analysis results.

{  Obviously the stellar mass determinations for single giants, based on isochrone fitting, are all, including our own, model dependent. We note, however a recent asteroseismic confirmation of our  determination (Paper III) of mass of HD 185351 (1.74$\pm$0.05 M$_{\odot}$) by \cite{Hjorringgaard2017} (1.58$\pm_{0.02}^{0.04}$ M$_{\odot}$), originally determined as 1.87$\pm$0.07 M$_{\odot}$ by \cite{Johnson2014} i.e. within 2.6-3.2$\sigma$ from the other two. In the case of  an independent asteroseismic mass determination for the same stars by  \cite{North2017}, i.e. 
1.77$\pm_{0.08}^{0.08}$ M$_{\odot}$,  our mass agrees even better, well within 1$\sigma$ of asteroseismic analysis. 
 These results allow us to consider our stellar mass estimates  as rather consistent with more precise determinations.

Also our stellar radii estimates, which are basically simple, appear to be very precise. In a recent study,  \cite{Baines2016} presented interferometrically measured radii for six stars from our sample. For four of these our radii agree with the interferometrically measured better than within 0.5$\sigma$ limit (HD 113226, HD 219615, HD 181276, HD 161797). For HD 188512 the radii agree within 1.07$\sigma$ and for HD 090537 within 1.57$\sigma$. }

The stellar age is the least precise parameter of all determined in our analysis, as the resulting Bayesian PDF for stellar age are generally flat  \citep{Adamczyk2016}.

{All results, including the adopted parallaxes, are presented in Tables 4 and 5.
A discussion is presented in Section \ref{PTPS_complete_sample}.
}

\section{Evolutionary status, subsample definition}\label{PTPS_sample}

\begin{figure}
\centering
 \includegraphics{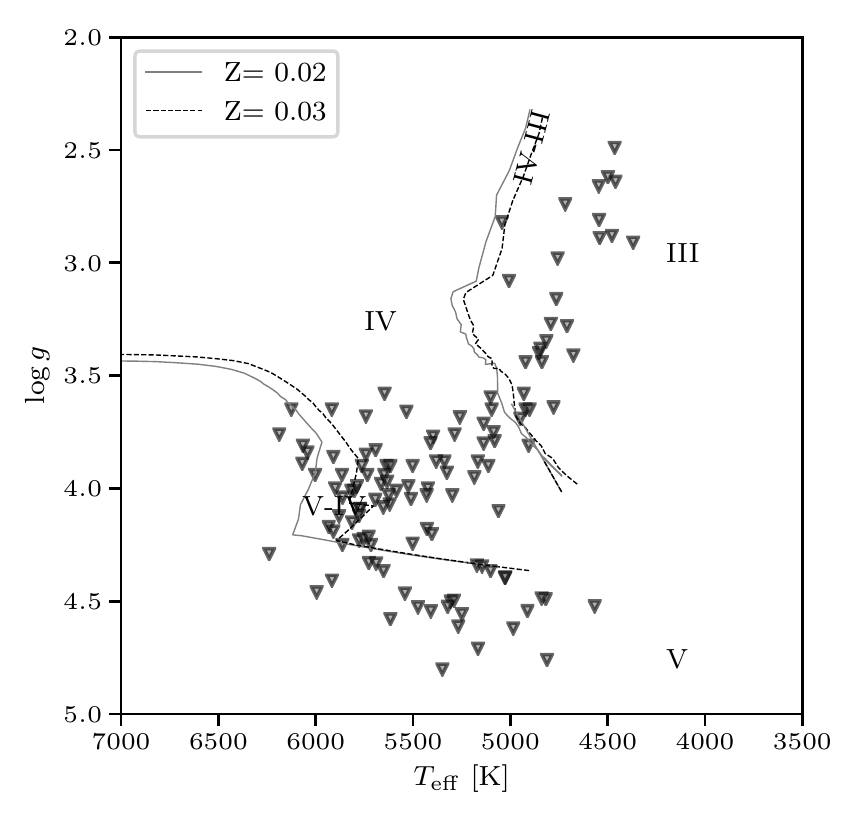}
\caption{Example of the HRD used to assign the luminosity class for PTPS stars. 
{ The lines joining the so-called critical points in evolutionary tracks correspond to two stages: the exhaustion of central hydrogen burning and climb to giant branch. These lines define three regions: V indicates dwarfs; IV indicates subgiants; and III indicates giants. See text of Section \ref{PTPS_sample} for detailed description.}}

\label{fig:hrch}
\end{figure}

The initial PTPS sample of red clump giants (Paper I) was successively extended with subgiants and giants (Paper III) and dwarfs (this work). 
The luminosity class (evolutionary stage)  was initially roughly known and the first, more solid, but still very approximate classification  presented in \cite{Adamow2010} comes from a photometric calibration.
 For simplicity, we initially decided to identify dwarfs stars with $\log g$ = 4.5$\pm$0.5, subgiants with $\log g$ = 3.5$\pm$0.5, giants with $\log g$ = 2.5$\pm$0.5, and bright giants with $\log g$ = 1.5$\pm$0.5. 

Currently, the detailed spectroscopic analysis facilitates the availability of  the luminosity class or evolutionary stage assessment, in principle, for every object. Indeed, for a vast majority of stars in our sample the evolutionary stage is obvious. However, given the uncertainties in the determined atmospheric or physical parameters at specific stages, such as dwarf to subgiant, or subgiant to giant borderlines, the evolutionary stage is still a bit problematic.

In an attempt to assign evolutionary status for all stars in the complete PTPS sample in a uniform and objective (as far as possible) way, we decided
to use the modified theoretical HRD,  a relation between $\log g$ and effective temperature. The analysis was based on the evolutionary tracks of \cite{Bressan2012}. We used the evolutionary tracks with metallicity values of Z = 0.0005, 0.001, 0.002, 0.004, 0.006, 0.008, 0.010, 0.014, 0.017, 0.020, 0.030, and 0.040, which approximately correspond to $\feh$ values, respectively, $\feh$ = -1.48, -1.18, -0.88, -0.58, -0.40, -0.28, -0.18, -0.04, 0.05, 0.12,
0.30, and 0.42, where Z = 0.0152 for Sun. We selected evolutionary tracks for stars with masses in the range between 0.75 and 3.80 M$_{\odot}$ and thus we constructed 12 various modified HRDs for different metallicities. { On each modified HRD we placed evolutionary tracks for two subsequent metallicities Z and we divided modified HRD  area into three regions:  dwarfs (V),  subgiants (IV), and   giants (III).} We defined these areas using the so-called critical points in evolutionary tracks, which correspond to two stages: exhaustion of central hydrogen burning and climb to giant branch. Three specific regions on the diagram are separated by lines  connecting  the same critical point for each mass (Figure \ref{fig:hrch}).

For example, on a single modified HRD with evolutionary tracks with Z=0.02 and Z=0.03, we placed  objects with 0.30 $>$ $\feh$ $\ge$ 0.12, while on the next we placed those of  Z=0.014 and Z=0.017 and  stars with -0.04 $\le$ $\feh$ $<$ 0.05. Therefore, we attributed objects to a given luminosity class  taking into account only the three most precisely derived atmospheric parameters: effective temperature, $\log g$, and metallicity.

 The procedure described above allowed us to assign the luminosity class to all PTPS stars, but it also revealed three groups of stars, which atmospheric and physical parameters were inconsistent with their apparent evolutionary stage.  For example the determined mass appeared  too low for assigned evolutionary stage. 

The most ambiguous group is  composed of six giants (HD 96692, TYC 3463-01145-1, TYC 3498-00634-1, HD 237903, BD+58 2015, and BD+63 639) situated at the bottom of the HRD, near the MS. Those  stars all appear to be very low mass (lower than 0.66 M$_{\odot}$) and with metallicity lower than solar ($\feh$ $<$ -0.31).

The second group of stars with inconsistent parameters and evolutionary stage contains seven subgiants (BD+42 170, TYC 3101-00586-1, TYC 3431-01221-1, BD+48 2320, TYC 3826-00664-1, BD-03 1821, and BD-12 2895) situated on left side of the MS.
 These are again stars with low mass (lower than 0.81 M$_{\odot}$) and metallicity mostly below solar.

 There is also an ambiguous star (HD 25532) situated on HRD between $\log L/L_{\odot}$ = 1 and $\log L/L_{\odot}$ = 2. According to our analysis this object is  a subgiant. However, its atmospheric parameters do not indicate clearly whether this star belongs to subgiants or is just beginning the ascent on the giant branch.

 Despite obtained atmospheric and physical parameters those 14 stars were rejected  from the dwarf sample and are not included in  the further analysis. These   
 are presented in Table 5. 
 
At this stage we also rejected from the PTPS sample two stars (HD 61606B, BD+46 1608) from the red clump giants sample, for which Gaia parallax based luminosity contradicted the evolutionary status of these stars as determined from atmospheric parameters.

In Table 5 we also placed, for completeness, the 13 stars of the 16 from Paper I for which inconsistent atmospheric parameters were derived. The three remaining stars were excluded from the PTPS sample based on CCF analysis.

\begin{figure}
\centering
 \includegraphics{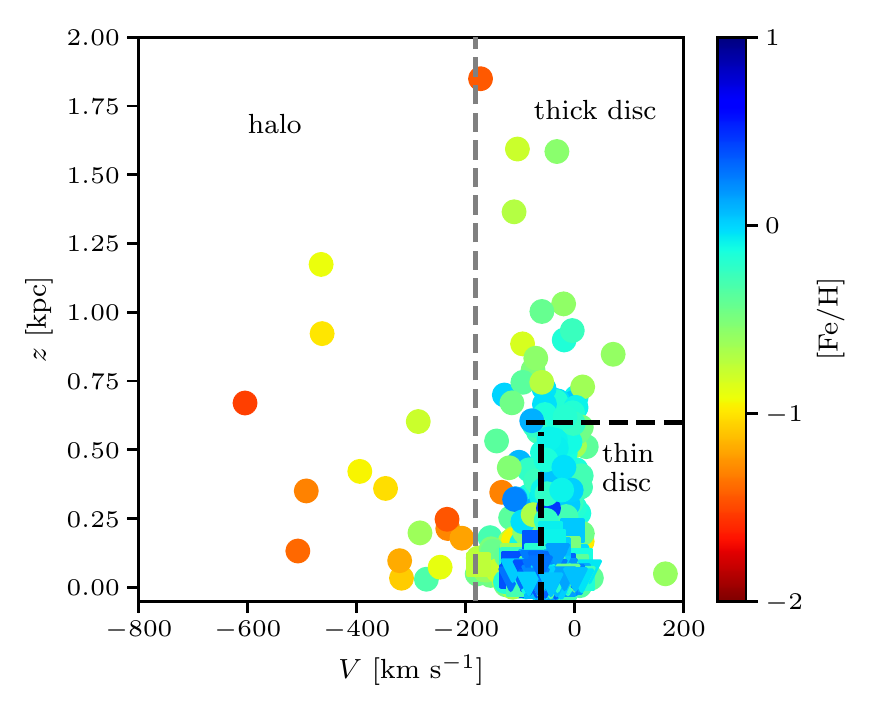}
\caption{Distribution of the distance from the Galactic plane (z) and the velocity in the direction of the rotation of the Galaxy (V) for the PTPS stars. Three stellar subsystems in the Galaxy are indicated according to \cite{Ibukiyama2002}. The stellar metallicity - [Fe/H] is colour coded.
}

\label{fig:galV}
\end{figure}

\section{Distribution of PTPS stars in the Galaxy}

\begin{figure*}
   \centering
   \includegraphics{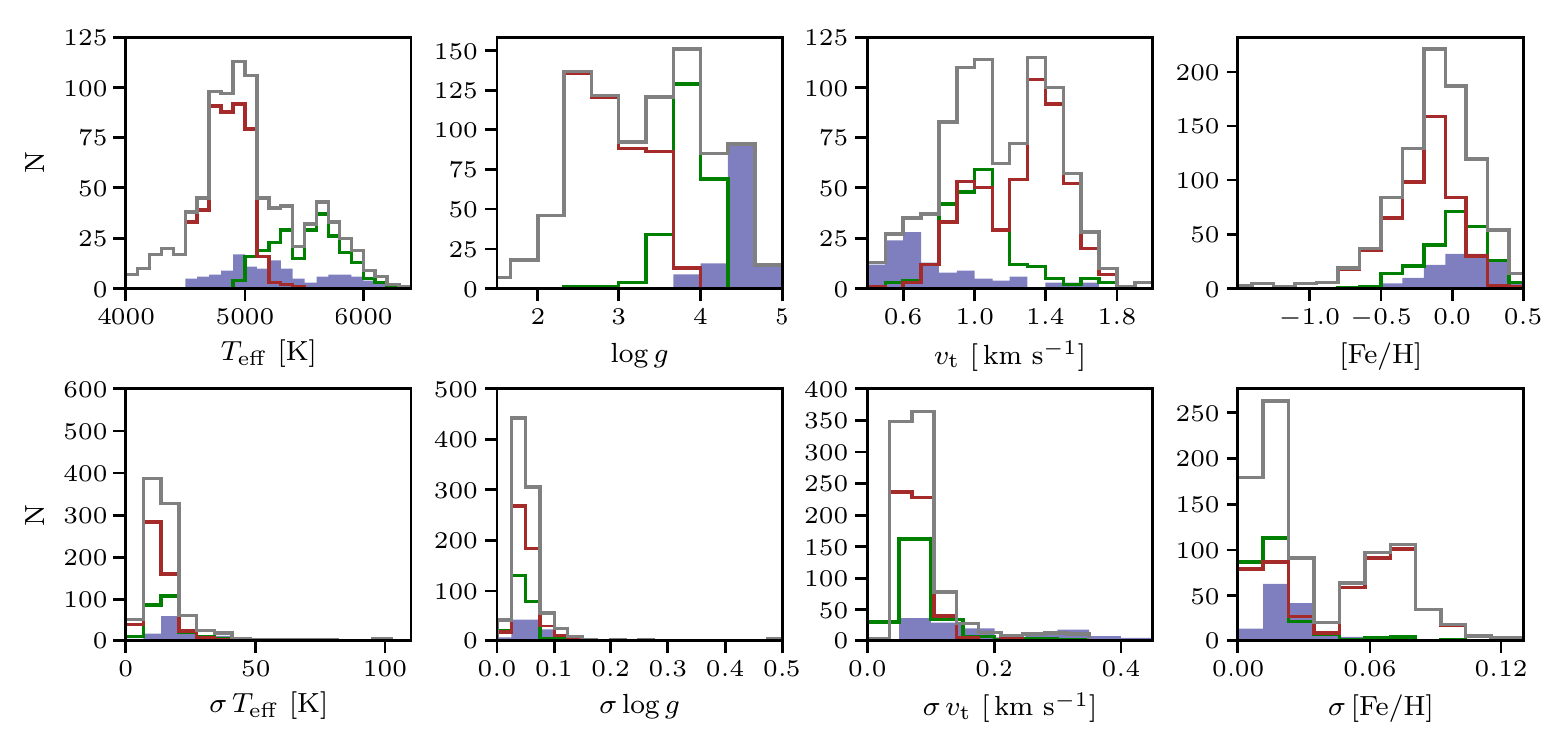}
   \caption{Histograms of the atmospheric parameters $\Teff$, $\log g$, $\vt$, and $\feh$ (upper panel) and their uncertainties (lower panel) for the complete PTPS sample of 
                 885 stars with uniform spectroscopic analysis. The subsamples, as defined in Section \ref{PTPS_sample}, are colour coded. The  blue area indicates dwarfs; the green line indicates subgiants; the brown line indicates giants; and the grey line indicates the whole sample. }
   \label{fig-histsigpar}
 \end{figure*}

\begin{figure*}
\centering
 \includegraphics{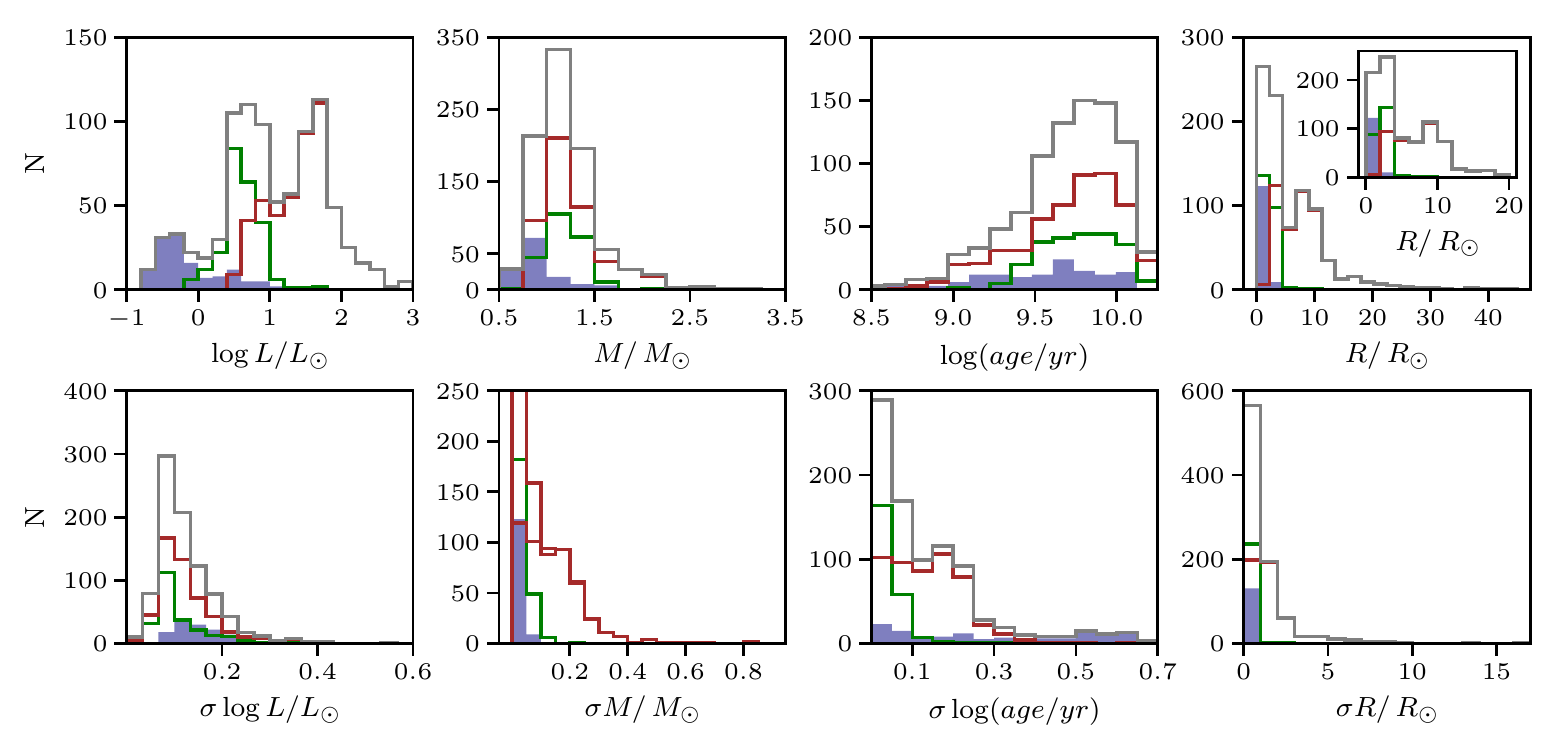}
\caption{Same as Figure \ref{fig-histsigpar} but for physical parameters stellar luminosity, mass, age, and radius.
}

\label{fig:allspar}
\end{figure*}

\begin{figure*}
   \centering
   \includegraphics{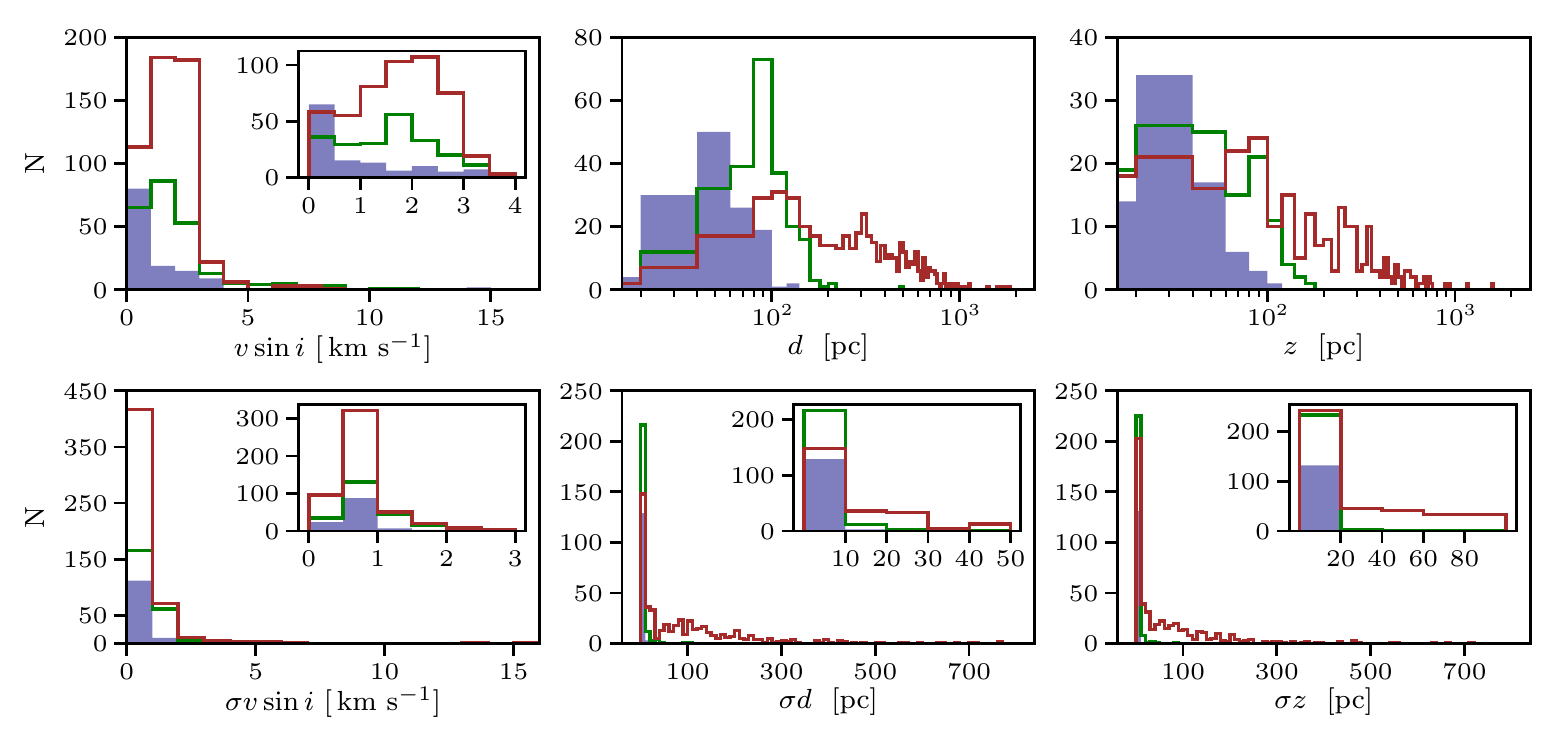}
      \caption{Same as Figure \ref{fig-histsigpar} but for projected rotational velocity, distance from the Sun (d), and  distance from the Galactic plane (z).}
   \label{fig-vrdz}
\end{figure*}

 To broaden our knowledge about the PTPS sample of stars we reviewed the distribution of PTPS stars in the Galaxy.
The absolute  RVs together with the Gaia \citep{Gaia2016} parallaxes and proper motions allow us to study kinematical properties of the PTPS sample. For all PTPS stars we determined the space velocity components (U, V, W)  following  the approach of \cite{Johnson1987}. Using the criteria described in \cite{Ibukiyama2002} we divided our sample into three Galactic populations: thin disc, thick disc, and halo. Figure \ref{fig:galV} presents distribution of PTPS stars in the Galaxy. 

It is clear that the PTPS stars sample an extensive volume of the Galaxy. Dwarfs and subgiants are mostly located within the thin disc. Giants, on the other hand, extend not only to thick disc but also a number of these are present in the Galactic halo.
The majority of our stars belong to the thin disc population (69$\%$), a rather large group belong to the thick disc (29$\%$), and another 2$\%$  belong to the Galactic halo.

In Figure  \ref{fig:galV} one can also see the relation between metallicity, in every population, with distance from the Galactic plane. 
The  stars from the Galactic halo present the lowest metallicity, which can be explained with the metal - deficient halo stars model \citep{Laird1988}. The other two populations are characterised by a much wider span of metallicity. Dwarfs and subgiants in thin and thick discs do not show a correlation between $\feh$ and $z$. In the case of giants in the thin Galactic disc no such a correlation  is present, but in the thick disc and halo a correlation between metallicity and distance from the Galactic plane is apparent. 
In Tables 4 and 5 we present kinematical properties of PTPS stars.

 \section{Complete PTPS sample}\label{PTPS_complete_sample}
 
Together with those already presented in Paper I and Paper III the stars presented here constitute the complete PTPS sample. After rejecting 111 objects with the CCF analysis (CCF variable, weak, or SB2), and another 29 stars for which inconsistent results were obtained, there are altogether 885 stars  in the complete PTPS sample. 

Atmospheric parameters (either new, determined in this paper, or those from papers I and III), new and updated masses, luminosities, radii, and ages,  as well as  $V$, $B-V$, original spectral type from Simbad, adopted parallax (from Gaia (G), Hipparcos (H), or Bayes (B) for stars without parallax or $\pi \leq$ 3), projected rotational velocity, $M_{V}$, $BC_{V}$, adopted luminosity class, absolute RV,  epoch of absolute RV observation,  distance from the Sun, $l$, $b$, $z$, $U$, $V$, $W,$ and resulting Galactic population  for these stars are presented in Tables 4 and 5. 

In Table 3 we present all stars rejected from the final sample after the CCF analysis and in Table 5  those rejected after further analysis. In Table 5, we list available data on these stars along with the reason they were rejected.
 
 In the complete PTPS sample of 885 stars we have 515 giants, 238 subgiants, and 132 dwarfs. The final HRD is presented in Figure \ref{fig:hr}. The atmospheric parameters for all these stars are presented in Figure \ref{fig-histsigpar}. The adopted masses, luminosities, radii, and ages are presented in Figure \ref{fig:allspar}.  In Figure \ref{fig-vrdz} we present rotational velocities (from \citealt{Adamow2014phd}), distances from the Sun, and distances from the Galactic plane. The distribution of the PTPS sample stars in the Galaxy is illustrated in Figure \ref{fig:galV}.}
 
 Below we present a short summary of properties of stars in all three luminosity class subsamples.

\subsection{Dwarfs}

The sample of dwarfs is  mainly composed of 101 objects presented for the first time in this paper. It also contains  30 stars from  Paper III and 1 object from Paper I classified here as dwarfs. The complete sample  of 132 dwarfs in PTPS sample presents a rather flat distribution of effective temperatures ranging from 4530 K to 6394 K. The median $\log g$ equals 4.52 and ranges from 3.65 to 4.80.  These stars are generally slightly more metal abundant than the Sun and have a median $\feh$ equal to 0.08 and ranging from -0.46 to 0.47. The microturbulence velocity in this sample ranges from 0.02 to 1.68 $\kms$ and have a median value of 0.65 $\kms$. These stars are slow rotators with projected rotational velocities below 4 $\kms$ and have  a median value of only 0.54 $\kms$.

The dwarfs are generally less luminous than the Sun and have a median $\log L/L_{\odot}$ equal to -0.25 and ranging from -0.71 to 1.35. These stars are also generally less massive than the Sun and have a median mass of 0.83 $\Msun$, ranging from 0.68 to 1.66. The dwarfs in our sample are also generally smaller than the Sun and have a median $R/\Rsun$ equal to 0.87 and ranging from 0.65 to 3.01. On average they are also younger, having a median age of 3.74 billion years and ranging from 0.11 to 13.20 Gyr.

The dwarfs in our sample belong to the solar neighbourhood within roughly 100 pc. Almost all of these objects also stay within 100 pc from the Galactic plane. Most of the dwarfs (92)  belong to the thin Galactic disc and 40 are located in the thick disc.

\subsection{Subgiants}

Our sample of 238 subgiants is  composed mainly of 186 stars presented in Paper III. The sample also contains a 44 objects from the sample studied in this paper and 8 stars from Paper I.  
Our sample is composed of stars with effective temperatures from  4906 K to 6215 K and has a clear peak in the distribution at median $\Teff$ of 5539K. Like the dwarfs, the subgiants present rather narrow range of $\log g$,  from 2.62 to 4.33, and have a clear peak at median $\log g$ of 3.86. Their metallicity is roughly solar and have a median value of $\feh$ equal to 0.04, ranging from -0.74 to 0.51. The microturbulence velocity ranges from 0.25 to 2.01 $\kms$ and have a median value of 1.01 $\kms$.  These stars are generally slow rotators with projected rotational velocity typically below 4 $\kms$, but in the case of 21 objects the rotational velocity is higher and may reach up to 11.8  $\kms$. In the case of 2 objects rotational velocities exceed 10 $\kms$ making them fast rotators.

\begin{figure*}
\centering
 \includegraphics{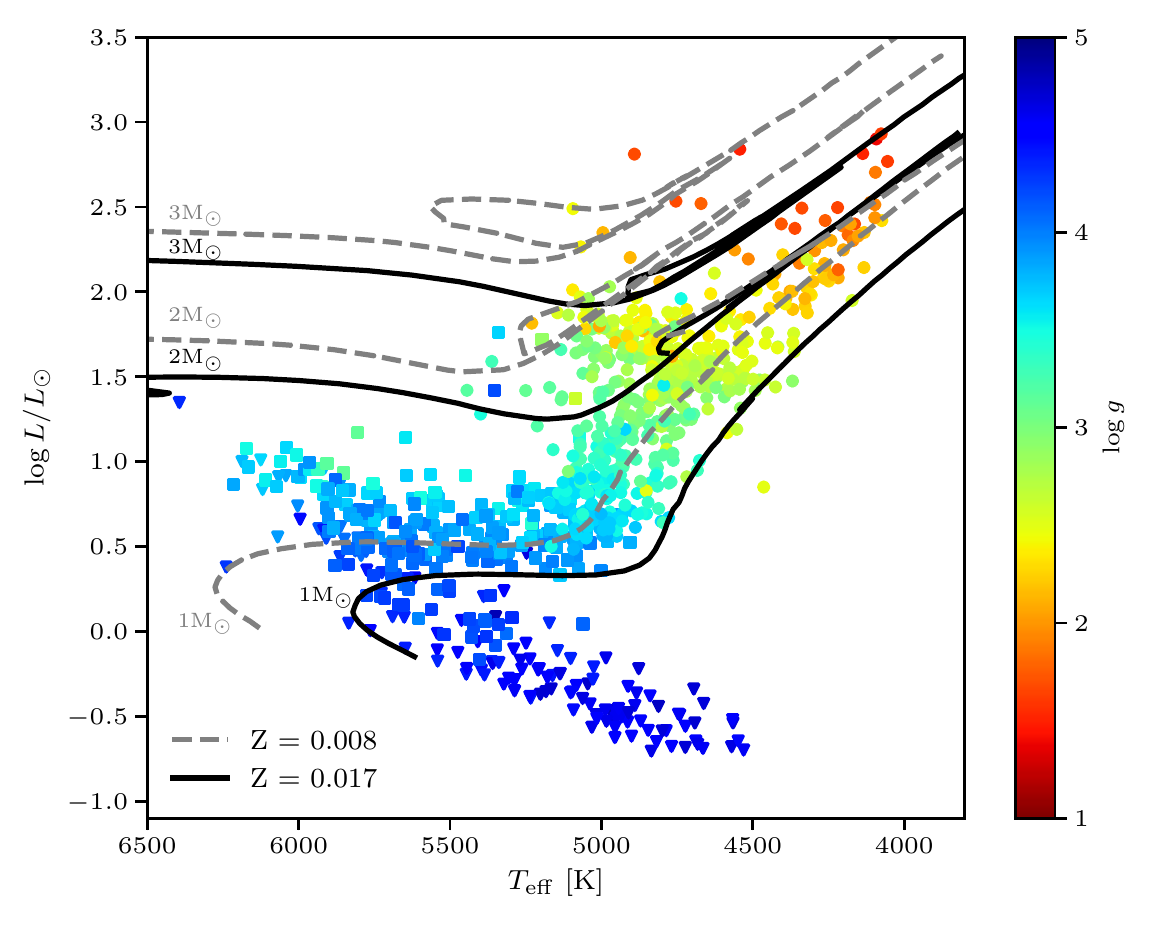}
\caption{The HRD  for all 885 stars from the PTPS sample with uniformly determined parameters.  Theoretical evolutionary tracks (Bertelli et al. 2008, 2009)  for stellar masses of 1, 2, and 3 M$_{\odot}$ and two
metallicities are presented as well.  Newly assigned luminosity class is coded with a shape: triangles indicate dwarfs; rectangles indicate subgiants; and circles indicate giants. The gravitational acceleration at stellar surface - log$\it{g}$ is colour coded.
}

\label{fig:hr}
\end{figure*}

On average these stars are more luminous than the Sun and have a median $\log L/L_{\odot}$ of 0.59, varying from -0.17 to 1.76. Their masses vary from 0.68 to 2.54 $\Msun$ and have a median value of 1.17 $\Msun$. These stars are typically larger than the Sun and have a median radius of 2.15 $\Rsun$, varying between 1.06 and 9.20 $\Rsun$. Their estimated ages vary between 0.63 and  13.42 Gyr and have a median of 6.04 Gyr.

The subgiants in our sample are located within roughly 200 pc from the Sun; most of these stars are located at a distances between 5 and 150 pc. Their distances from the Galactic plane are also limited to about 150 pc. Hence vast majority of the subgiants (163) belong to the thin disc and only 75 reside in the thick disc.

\subsection{Giants}{\label{PTPS_giants}}

The sample of  giants consists mainly of objects  from Paper I (320 stars). This sample also contains  184 stars presented in Paper III and 11 objects from the sample presented here for the first time. This is the  largest PTPS sample of 515 stars and it contains giants in a range of effective temperatures from 4055 K to 5444 K and have a clear peak at a median value of  4820 K. The surface gravitational accelerations $\log g$ also belong to a wide range from 1.39 to 3.79 and have a peak of the distribution at a median value around 2.77. These stars are on average much less metal abundant than the Sun and have a median $\feh$ of -0.18, ranging from -1.52 to 0.45. The microturbulence velocity in our giants varies between 0.43 to 2.04 $\kms$ and have a median value of 1.34 $\kms$. These stars are generally slow rotators; the vast majority show projected rotational velocities below 4 $\kms$. However, in the case of 14 objects the rotational velocity is slightly higher (up to  8.2 $\kms$).

Our giants display a wide range of luminosities between $\log L/L_{\odot}$ 0.48 and  2.93 and have a median value of 1.52. Also their masses belong to a wide range from 0.48 to 2.93 $\Msun$ and have a median of 1.19 $\Msun$. When it comes to physical dimensions their radii range from 2.12 to 44.52 $\Rsun$ and have a median of 8.27 $\Rsun$. The wide range of masses leads to wide range of estimated ages, between 0.32 to 13.43 Gyr and a median of 5.74 Gyr.

The giants in our sample reside at a wide range of distances from the Sun, reaching up to nearly 2 kpc. Most of them stay within 1 kpc, and the distribution of distances is rather flat in a wide range of 100-400 pc.  The distribution of distances from the Galactic plane is similar; most of giants are located   below $z$=500 pc. The large distances to these stars result in many of the giants (153) already being members of the thick disc and 16 being located in the Galactic halo.

\section{Summary and  conclusions}\label{conclusions}

 We presented the complete PTPS sample of 885 stars. For 156  stars, we presented new atmospheric parameters, masses luminosities, radii, and ages.
For 135 stars these are the first determinations.

For another 451 stars we presented updated masses, luminosities, radii, and ages based on new Gaia parallaxes. We also presented a list of 140 stars that were rejected from the original PTPS sample for various reasons, as explained in the text.

The complete PTPS sample of 885 stars is composed of  515 giants, 238 subgiants, and 132 dwarfs.
For most of the dwarfs (114 stars) Gaia parallaxes were available, for 13 we accepted Hipparcos parallaxes, and only for 5 we were forced to use luminosities derived from Bayesian analysis.
In the case of subgiants, we also mostly base our analysis on solid parallaxes either from Gaia (179 stars) or Hipparcos (47), and only for 12 subgiants stellar luminosities are derived from Bayesian analysis.
For giants, however, in spite of Gaia parallaxes first release, we still mostly based our analysis on Bayesian luminosity determinations, which are the only available values for 281 stars. For 158 giants we used Gaia parallaxes, and for 76 those from Hipparcos.

In general the masses of PTPS  stars  range from 0.68 to 3.21 M$_{\odot}$ with a median at 1.14  M$_{\odot}$.
The vast majority of stars have masses between 1.00 and 1.25 M$_{\odot}$. However, there are also   114 stars that have masses over 1.5 M$_{\odot}$, 30 of which have masses  above two solar masses.

Figure \ref{fig:hr} presents the HRD with the new and updated luminosities for all 885 PTPS stars.

\begin{figure*}
\centering
 \includegraphics{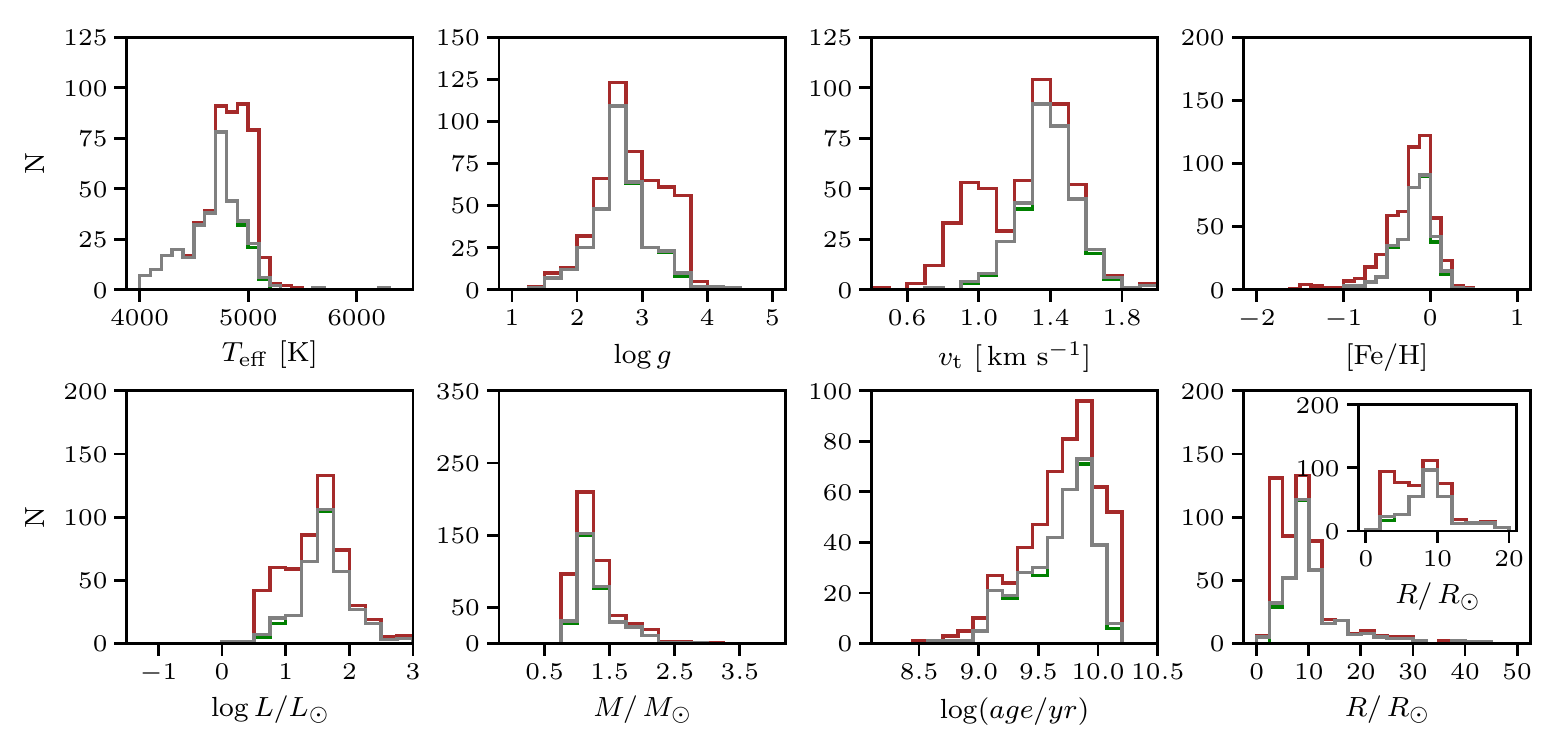}
\caption{Evolution of the PTPS giants subsample definition. Histograms of atmospheric parameters: $\Teff$, $\log g$, $\vt$ and $\feh$ (upper panel) as well as luminosities, masses, ages, and radii (lower panel) for the complete PTPS sample of 
                 red giant stars with uniform spectroscopic analysis are shown. The evolution of the subsample definition is colour coded:  
                 the red line indicates the final red giants sample of 515 stars as defined in this paper; 
                 the green line indicates giants from Paper I only { (320 red clump giants)}; and the 
                 gray line  indicates all 348 stars from Paper I.
}

\label{fig:newgiants}
\end{figure*}

{

We find it important to note here that with this paper we redefine the subsamples of PTPS. As various simple criteria were applied previously to 
select objects for PTPS or later on to 
classify PTPS stars into luminosity classes (see Section \ref{PTPS_sample}) the final subsamples as defined in this paper all contain a mixture of stars from Paper I, Paper III, and this paper (see Section \ref{PTPS_complete_sample}).

Obviously the subsample of dwarfs did not evolve much as its is presented here for the first time. The only alteration of the original subsample comes from additional 31 stars from Paper I and Paper III 
and the removal of stars not classified as dwarfs from the sample.

 The subsample of subgiants was originally presented in Paper III, together with numerous red giants from the red giant branch. The subsample was  cleaned from the red giants and  extended with 52 stars from Paper I and this paper. Consequently the final subgiant subsample consists of stars with more uniform parameters. 

 Historically the first subsample of PTPS stars, giants, was at first composed of presumably red clump giants presented in Paper I. This subsample was extended with 184 giants from the red giants branch presented in Paper III and 11 giants from the sample of stars studied in this paper. This is the largest subsample within PTPS (515 stars). The evolution of giants subsample definition is illustrated in Figure
 \ref{fig:newgiants}. This Figure shows how the sample evolved from a red clump giant sample into a general red giant sample.
This sample may be subdivided into clump giants (320 giants from Paper I only) and red giants (the complete subsample)  with slightly different characteristics (see Figure \ref{fig:newgiants}).  

This PTPS giant sample is the most intensively studied  sample so far. On average 10 epochs of HET/HRS RV are available per target. 
 No surprise then that all but one (BD+14 4559, \citealt{Niedzielski2009}) planetary systems delivered by PTPS are hosted by giants \citep{Niedzielski2007,Niedzielski2009,Niedzielski2009a,Adamow2012,Gettel2012a,Gettel2012b,Nowak2013,Niedzielski2015}. 
 Moreover this sample was recently more intensively explored within TAPAS project with Harps-N at TNG \citep{tapas1,tapas3,tapas4,tapas5}.

 In total  21 planetary systems were already discovered in the red giant sample of 515 stars or 20 in the red clump sample of 320 stars. With only  $\sim$4$\%$ of stars with planets the search for planets around giants within PTPS is yet far from complete. The population of stars with planets around giants already detected within PTPS is however already similar to those presented by other similar projects (15 out of 373 stars, $\sim$ 4$\%$ in the   Lick K-giant Survey: \citealt{Reffert2015}; 
 10 out of 166, $\sim$ 6$\%$ in EXPRESS: \citep{Jones2016}; 
 or 11 out of 164, $\sim$ 7$\%$ in  the Pan-Pacific Planet Search: \citealt{Wittenmyer2017}).  We are now in position to present statistical properties of planet hosting giants based on our own data, which will be the topic of a subsequent paper.

\begin{acknowledgements}

We thank the HET resident astronomers and 
telescope operators for their continuous support. 
AN and BD-S 
were supported by the Polish National Science Centre
grant no. UMO-2015/19/B/ST9/02937.
MA acknowledges the Mobility+III fellowship from the Polish Ministry of Science
and Higher Education. 
The HET is a joint project of the University of Texas 
at Austin, the Pennsylvania State University, Stanford University, 
Ludwig-Maximilians-Universit\"at M\"unchen, and Georg-August-Universit\"at 
G\"ottingen. The Hobby-Eberly Telescope is named in honor of its principal benefactors, William P. 
Hobby and Robert E. Eberly. The Center for Exoplanets and Habitable Worlds is 
supported by the Pennsylvania State University, the Eberly College of Science, 
and the Pennsylvania Space Grant Consortium. This research has made extensive 
use of the SIMBAD database, operated at CDS (Strasbourg, France) and NASA's 
Astrophysics Data System Bibliographic Services.

This work has made use of data from the European Space Agency (ESA)
mission {\it Gaia} (\url{https://www.cosmos.esa.int/gaia}), processed by
the {\it Gaia} Data Processing and Analysis Consortium (DPAC;
\url{https://www.cosmos.esa.int/web/gaia/dpac/consortium}). Funding
for the DPAC has been provided by national institutions, in particular
the institutions participating in the {\it Gaia} Multilateral Agreement.

We thank the anonymous referee for helping us to improve the manuscript substantially.
\end{acknowledgements}

\bibliographystyle{aa} 
\bibliography{PTPS_dwarfs_fin} 

\end{document}